\documentclass[amsmath,twocolumn]{astronomycomm}

\begin{document}

\title{Dark Matter Particle Explorer: Ten-Year On-Orbit Performance and Scientific Outcomes
}

\author[orcid=0000-0002-5778-8228,gname=Given,sname=Family]{Jian-Hua Guo}
\affiliation{Purple Mountain Observatory, Chinese Academy of Sciences,
Nanjing 210023, China}
\affiliation{School of Astronomy and Space Science, University of Science and
Technology of China, Hefei 230026, China}
\email{jhguo@pmo.ac.cn}

\author[orcid=0000-0002-1345-092X,gname=Given,sname=Family]{Chuan Yue}
\affiliation{Purple Mountain Observatory, Chinese Academy of Sciences,
Nanjing 210023, China}
\affiliation{School of Astronomy and Space Science, University of Science and
Technology of China, Hefei 230026, China}
\email{yuechuan@pmo.ac.cn}

\author[orcid=0000-0002-0785-6827,gname=Given,sname=Family]{Yun-Long Zhang}
\affiliation{State Key Laboratory of Particle Detection and Electronics, University of Science and Technology of China, Hefei 230026, China}
\affiliation{Department of Modern Physics, University of Science and Technology of China, Hefei 230026, China}
\email{ylzhang@ustc.edu.cn}

\author[orcid=0000-0002-8966-6911,gname=Given,sname=Family]{Yi-Zhong Fan}
\affiliation{Purple Mountain Observatory, Chinese Academy of Sciences,
Nanjing 210023, China}
\affiliation{School of Astronomy and Space Science, University of Science and
Technology of China, Hefei 230026, China}
\email{yzfan@pmo.ac.cn}

\author[orcid=0000-0003-0066-8660,gname=Given,sname=Family]{Jin Chang}
\affiliation{Purple Mountain Observatory, Chinese Academy of Sciences,
Nanjing 210023, China}
\affiliation{School of Astronomy and Space Science, University of Science and
Technology of China, Hefei 230026, China}
\email[show]{chang@pmo.ac.cn}

\begin{abstract}

The Dark Matter Particle Explorer (DAMPE, also known as ``Wukong'') is the first scientific space mission within the framework of the Strategic Pioneer Program on Space Science of the Chinese Academy of Sciences. The main scientific objectives of DAMPE include indirect detection of dark matter, cosmic-ray physics, and high energy gamma-ray astronomy. Thanks to its large geometrical factor and thick imaging calorimeter, DAMPE could effectively detect cosmic-ray electrons (including positrons) and gamma-rays up to 10~TeV, and nuclei up to PeV. Since launched on 17 December 2015 into a Sun-synchronous orbit, DAMPE has been smoothly operating on orbit for more than ten years. Here we present an overview of instrument, on-orbit status, and highlight scientific results of DAMPE, including the measurements of the all-electron energy spectrum and the spectra of protons and other nuclei, and high energy gamma-ray line search.

\end{abstract}


\section{Introduction}
\label{sec:intro}
The origin of cosmic rays (CRs) and the particle nature of dark matter (DM) remain two fundamental open questions in astroparticle physics. CRs are generally thought to be accelerated in astrophysical environments such as supernova remnants \citep{Blasi:2013rva}, but their sources, acceleration mechanisms, and subsequent propagation through the Galaxy are not yet fully understood \citep{Strong:2007nh}. At the same time, the existence of DM is supported by a wide range of gravitational evidence, yet its microscopic identity remains unknown \citep{PDG:2024cfk}. If DM particles annihilate or decay, they may produce detectable secondary particles—including positrons, antiprotons, and gamma rays—making cosmic-ray observations a potential avenue for indirect searches \citep{Bertone:2004pz}. Precise measurements of the energy spectra, arrival directions and temporal variations of various CR species are essential to the understanding of CR origin and propagation, and could act as probes for indirect DM detection \citep{2005PhR...405..279B,2010ARA&A..48..495F,2010IJMPD..19.2011F,2015ARA&A..53..199G}.  

The Dark Matter Particle Explorer (DAMPE, also known as ``Wukong'' in China; \citet{Chang:2014}), is a satellite-borne, calorimetric particle detection mission, which is dedicated to studies of CR physics and the indirect detection of DM via precise observations of high energy CRs and gamma-rays  \citep{DAMPE:2017cev}. The satellite was successfully launched into a Sun-synchronous orbit at an altitude of about 500~km and an inclination of about $97^\circ$ on December 17, 2015 from the Jiuquan launch base. It is the first scientific space mission within the framework of the Strategic Pioneer Program on Space Science of the Chinese Academy of Sciences. 

The DAMPE detector is optimized to precisely measure CR electrons (including positrons) and gamma-rays with powerful hadron-rejection capability and excellent energy resolution from GeV to 10 TeV, for sensitive search of indirect signatures from DM particles \citep{DAMPE:2017cev}. DAMPE is also capable of measuring the energy spectra and the relative abundances of CR nuclei up to unprecedented PeV scale in space, thanks to its unique large geometrical factor and excellent charge resolution \citep{DAMPE:2017cev}. The main scientific objectives of the DAMPE mission include: (1) understanding the origin, acceleration and propagation of Galactic CRs; (2) probing the nature of DM; and (3) studying the high energy gamma-ray emissions from Galactic and extragalactic sources.

In this paper, an overview of the DAMPE instrument and its on-orbit status is provided, and the highlight scientific results of DAMPE including the spectra of all-electrons, protons and nuclei, and high energy gamma-ray line search are reported and discussed.

\section{DAMPE instrument}
\label{sec:mission}

\subsection{Detector configuration}
\label{sec:detector}

\begin{figure}[tbp]
\centering
\includegraphics[width=0.47\textwidth]{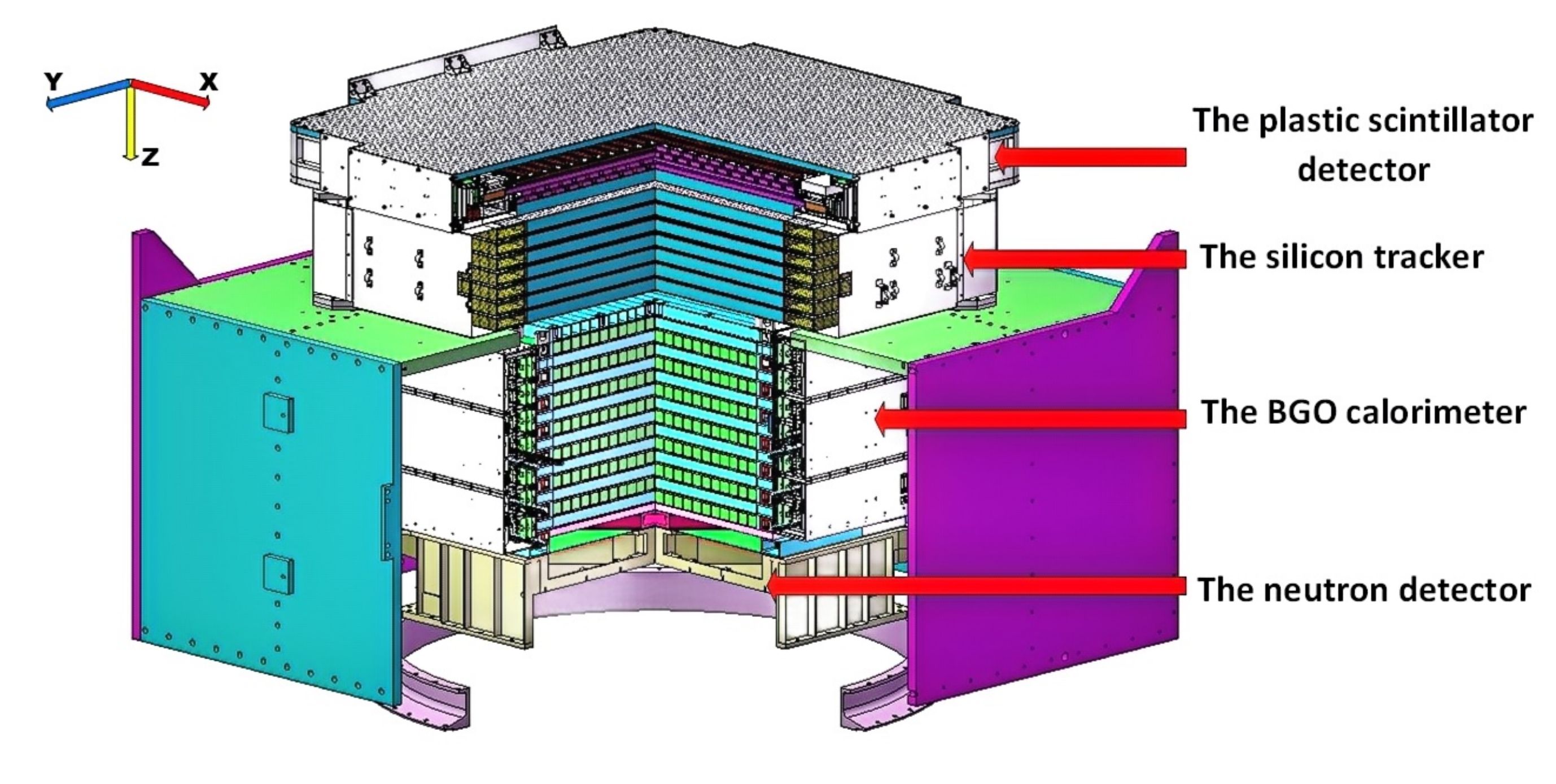}
\caption{Schematic view of the DAMPE instrument \citep{DAMPE:2017cev}, showing the four
subdetectors from top to bottom. The coordinate axes of the payload are indicated at upper left.}
\label{fig:payload}
\end{figure}

The DAMPE instrument is composed of four sub-detectors: a Plastic Scintillator strip Detector (PSD), a Silicon-Tungsten tracker-converter (STK), a Bismuth-Germanium Oxide (BGO) imaging calorimeter, and a NeUtron Detector (NUD) \citep{DAMPE:2017cev}. The schematic view of the DAMPE instrument is shown in Fig.~\ref{fig:payload}. As illustrated, four subdetectors, placed from top to bottom, work together to measure particle-ID, track and energy information of the incident particles. 

The PSD at the top provides the primary
charge measurement through the specific ionization energy loss of the incident charged particle and acts as an anti-coincidence veto for gamma-ray identification \citep{Yu:2017dpa}. It consists of 82 organic plastic scintillator bars arranged
in X \& Y planes, each with a double layer configuration, covering an
active area of 82.5 $\times$ 82.5 $cm^{2}$. The
scintillation light generated in each bar is collected
and read out by one photomultiplier tube (PMT) at each end. The PSD and related electronics are designed to measure the absolute charge values of incident particles over a large dynamic range from protons ($Z=1$) up to heavy nuclei ($Z>30$).

The STK is designed to measure the arriving trajectory of the incident particle with a high angular resolution \citep{Azzarello:2016trx}. It also supplies an independent charge measurement of light nuclei ($Z \leq 8$) via the ionization signal.
It is composed of 6 position-sensitive double (X \& Y) planes of silicon strips with a total area of about 7 $m^{2}$. The STK strips are 48 µm wide and 93 mm long, with a pitch of 121 µm, enabling a position resolution better than 80 µm \citep{Tykhonov:2017yae}. Multiple thin tungsten plates with a total thickness of about one radiation length are inserted into the first three silicon planes in order to enhance the photon conversion to electron-positron pairs. 

The BGO calorimeter is designed to precisely measure the total energy of the incident particle and distinguish between electromagnetic and hadronic particles by imaging the 3D profile of the shower development \citep{Zhang:2016xkz}.
It is composed of 308 BGO scintillation crystals arranged hodoscopically in 14 layers, covering an
active area of 30 $\times$ 30 $cm^{2}$.
The scintillation light
signal of each BGO crystal is collected by two photomultiplier tubes (PMTs) at both ends and each PMT is read out at three dynodes (2, 5, and 8), corresponding to low,
middle, and high gain, respectively. The two PMTs are coupled to the crystal with two optical filters whose attenuation ratios differ by a factor of 5, which allows each crystal to measure energy deposit from 0.5 MIPs (Minimum Ionizing Particles)
to $10^{5}$ MIPs \citep{zhangYL:2012}.
The total thickness of BGO calorimeter is equivalent to about 32 radiation lengths ($X_{0}$) or 1.6 proton interaction lengths ($\lambda_{I}$), ensuring effective containment for electromagnetic showers up to 10 TeV.

The NUD at the bottom is to perform electron/hadron discrimination by measuring the secondary neutrons mainly produced in the hadronic shower \citep{Huang:2020skz}. It is composed of four identical blocks of boron-loaded plastic scintillators with active area of 30 $\times$ 30 $cm^{2}$. Above 1 TeV, NUD provides independent proton rejection power by a factor of $\sim$2 on the basis of BGO calorimeter \citep{Huang:2020skz}. 

\begin{figure}[tbp]
\centering
\includegraphics[width=0.47\textwidth]{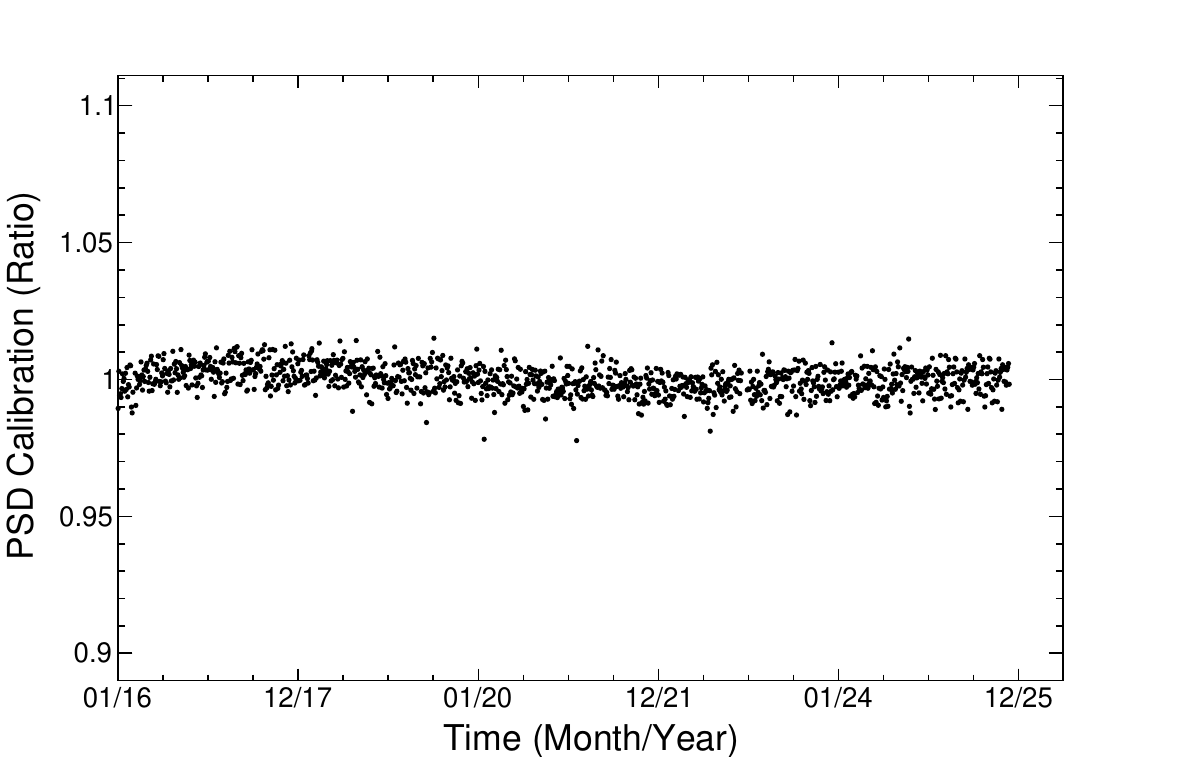}
\includegraphics[width=0.47\textwidth]{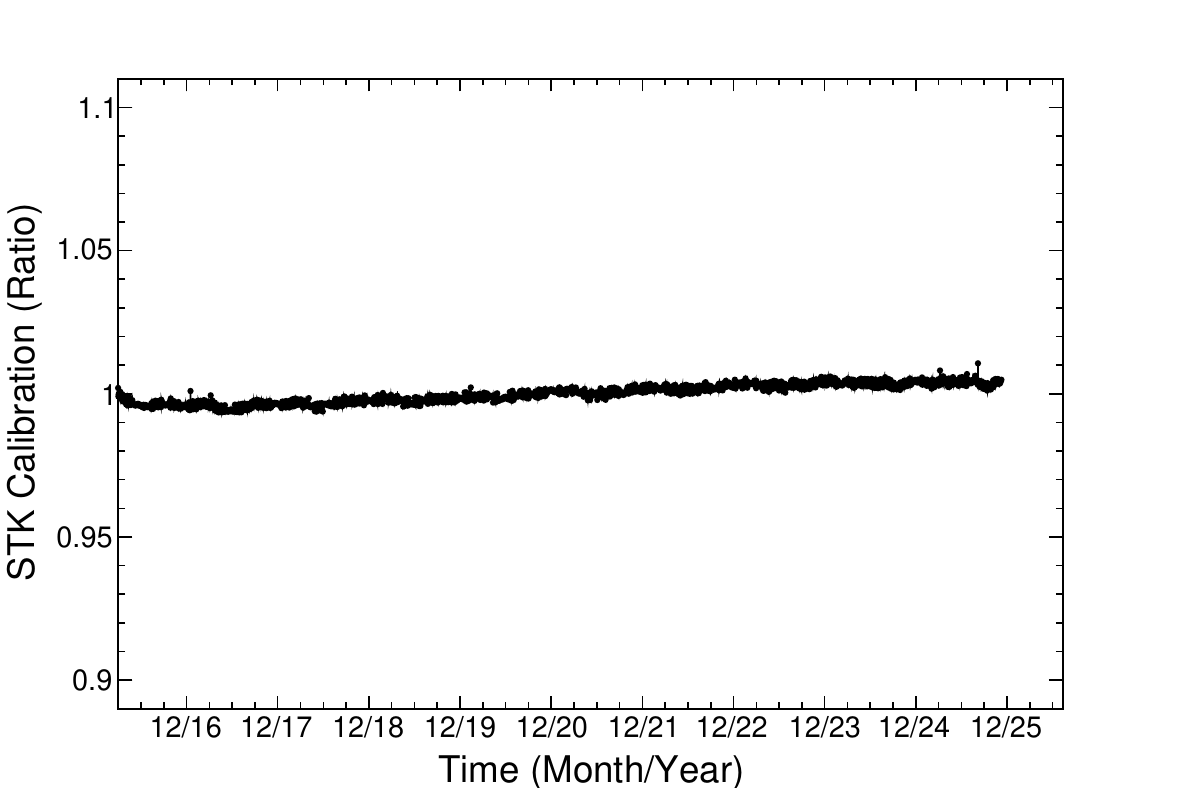}
\includegraphics[width=0.47\textwidth]{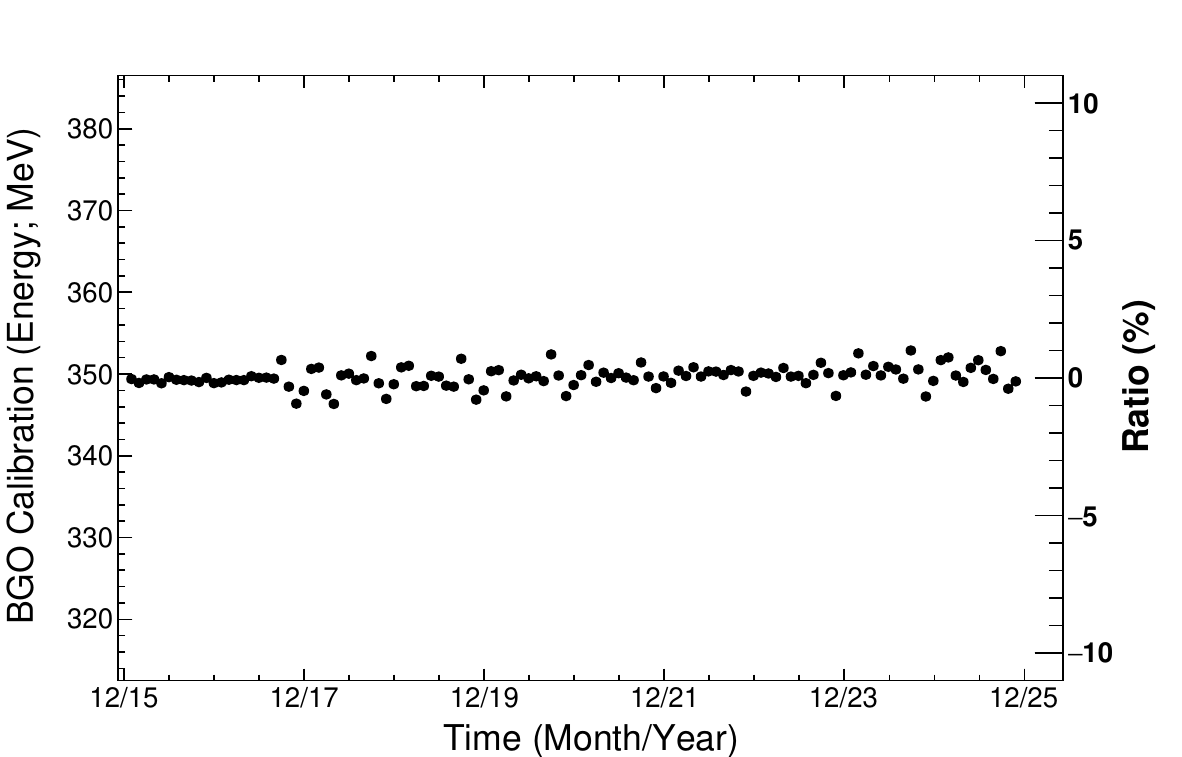}
\includegraphics[width=0.47\textwidth]{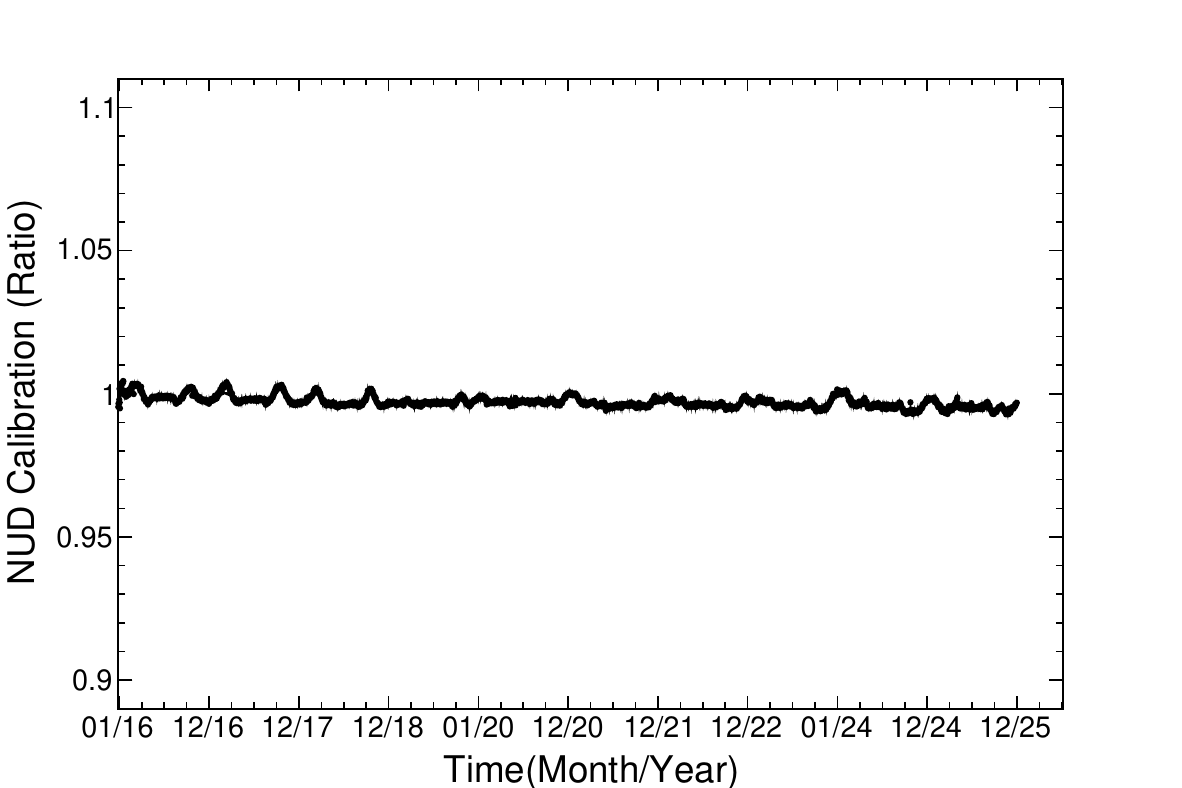}
\caption{Long-term stabilities of four subdetectors, from top to bottom, PSD: the average PMT gain, STK: the average electronic noise, BGO: the total energy measurement of Helium MIPs, and NUD: the average pedestal noise.}
\label{fig:orbit}
\end{figure}

\subsection{On-orbit operation}
\label{sec:operation}
Over the past decade, the DAMPE satellite has operated stably in space under the sky-survey mode. 
The daily event rate of DAMPE is about 5 million, and more than 18 billion events were recorded by the end of 2025.
Dedicated on-orbit calibration shows that the DAMPE instrument has maintained a good long-term performance \citep{DAMPE:2019lxv,Li:2024xsw}, which enables precise measurements of the physical quantities of each event. The long-term stabilities of four sub-detectors are illustrated in Fig.~\ref{fig:orbit}.

The charge measurement performance of PSD is tested by on-orbit CR nuclei events, and the charge resolution is about 0.18$e$ for Z=6 (Carbon) and 0.30$e$ for Z=26 (Iron) \citep{Ma:2018brb,Dong:2018qof}.
The angular resolution of the STK is well calibrated using bright gamma-ray sources, which is better than $0.2^{\circ}$ for energies higher than 10 GeV \citep{Duan:2024vzr}. The considerable thickness of BGO calorimeter, i.e. 32 $X_{0}$, achieves an excellent energy resolution close to 1\% for energies above 100 GeV \citep{YueC:2017}. 
The absolute energy scale of DAMPE is determined with the geomagnetic
cutoff of the all-electron spectrum, which results in a small
bias of about 1.3\% \citep{Zang:2025erm}. The linearity of the energy measurement is validated with the electron beam test and laser test up to a few TeV \citep{Zhang:2016xkz,Zhao:2022wfw}. The energy measurement for CR nuclei is extended to PeV scale after the BGO read-out saturation is corrected effectively \citep{Yue:2020hmj,Serpolla:2025vrj}.

Extensive Monte Carlo (MC) simulations are carried out to obtain the instrument response of DAMPE based on GEANT4 toolkit \citep{GEANT4:2002zbu}. The energy response of MC simulations is corrected by including the Birks' quenching effect for the ionization energy deposit in the BGO calorimeter \citep{Chen:2023koi}. An independent analysis based on FLUKA \citep{Bohlen:2014buj} is also performed to evaluate the uncertainties from the interaction model.

\section{Scientific results}
\label{sec:results}

\subsection{All-electron spectrum}
\label{sec:CRE}

The DAMPE instrument is optimized for the precise measurement of the all-electron (electron plus positron) spectrum from a few GeV to 10
TeV with a geometrical factor of $\sim$ 0.3 $m^{2}sr$. The 32 $X_{0}$ thick BGO calorimeter ensures a full containment of electron showers up to 10 TeV, with an excellent energy resolution (better than 1.2\% above 100 GeV). The different shower development features of electromagnetic and hadronic particles are exploited to separate electrons from protons, based on the 3D imaging measurement of BGO calorimeter. This electron/proton discrimination method is capable of rejecting $>$99.99\% of protons while keeping 90\% of electrons and positrons. The residual contamination of the proton background is estimated to be less than 3\% up to 1~TeV. 

\begin{figure}[tbp]
\centering
\includegraphics[width=0.47\textwidth]{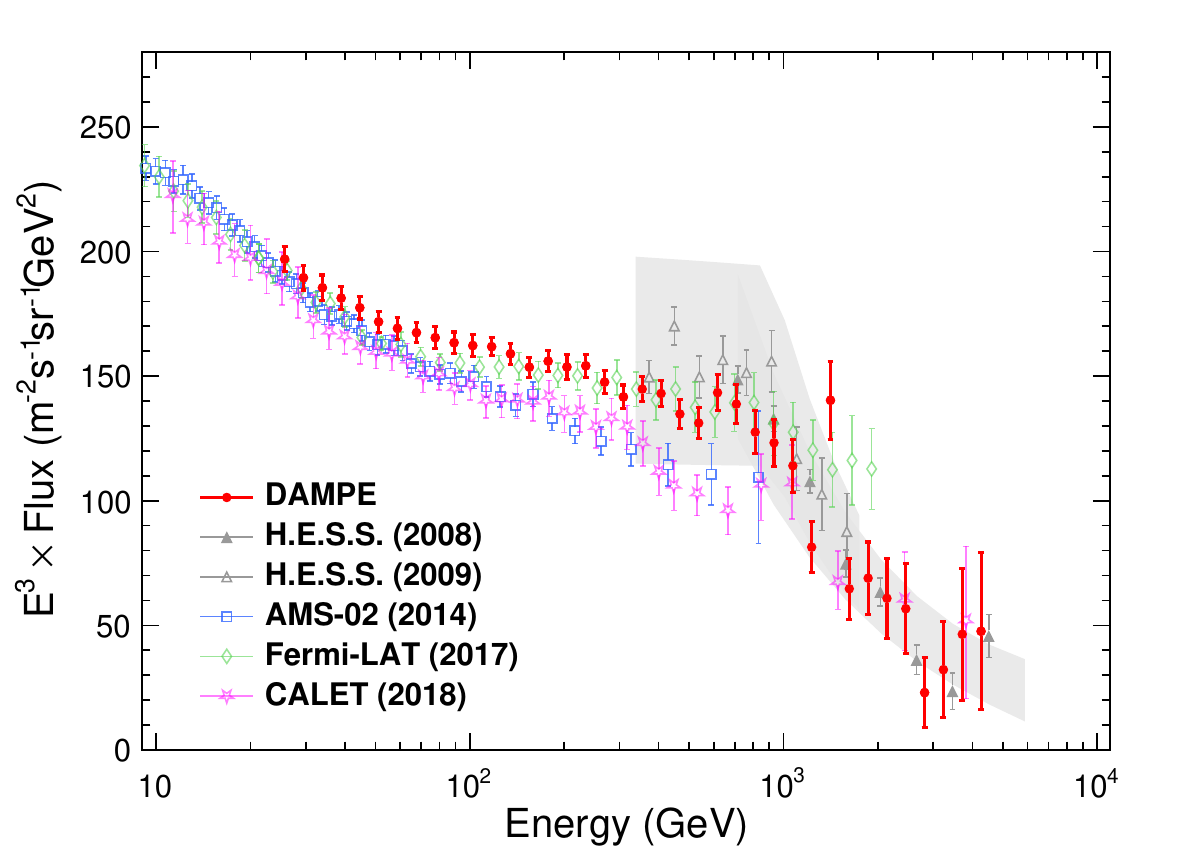}
\caption{The all-electron spectrum (multiplied by $E^3$) measured by DAMPE \citep{DAMPE:2017fbg}. The red dashed line represents the best fit of DAMPE data with a smoothly broken power-law model in the range from 55 GeV to 2.63 TeV. Also shown are the direct measurements from the space-borne experiments AMS-02 \citep{AMS:2014gdf}, Fermi-LAT \citep{Fermi-LAT:2017bpc} and CALET \citep{Adriani:2018ktz}, and the indirect measurement by the H.E.S.S.~Collaboration \citep{HESS:2008ibn,HESS:2009chc}, whose grey band represents its systematic errors apart from an approximately 15\% energy-scale uncertainty. The error bars of the DAMPE, AMS-02 and Fermi-LAT points include both statistical and systematic uncertainties added in quadrature.}
\label{fig:electron}
\end{figure}

The all-electron spectrum measured with DAMPE is shown in Fig.~\ref{fig:electron}, based
on the first 530~days of on-orbit operation \citep{DAMPE:2017fbg}. The spectrum integrates about 1.5~million electron and positron events from 24~GeV to 4.6~TeV. The systematic uncertainties of the flux measurement, with dominant contributions from the background subtraction and the effective acceptance, are described in detail in \citep{DAMPE:2017fbg}. The DAMPE spectrum is compared with previously published results from the space-borne experiments AMS-02 \citep{AMS:2014gdf} and Fermi-LAT \citep{Fermi-LAT:2017bpc}, as well as the ground-based experiment of the H.E.S.S. Collaboration \citep{HESS:2008ibn,HESS:2009chc}. 

Compared with previous experiments, DAMPE has significantly improved the accuracy of the all-electron spectrum, especially in the energy range above TeV. The DAMPE data in the energy range from 55~GeV to 2.63~TeV fit better to a smoothly broken power-law model than to
a single power-law model, suggesting a spectral break at E $\sim$ 0.9~TeV. The significance of the break is about 6.6$\sigma$. The break has been corroborated by updated AMS-02 \citep{AMS:2021nhj} and CALET \citep{CALET:2023emo} measurements. The precise
measurement of the all-electron spectrum by DAMPE, together with the positron measurements by AMS-02, has enhanced our understanding of the spectral structures related to nearby pulsars or dark matter \citep{Fan:2017sor,Yuan:2018rys,Zu:2017dzm}. With accumulated statistics and improved analysis method, DAMPE is expected to precisely measure the all-electron spectrum up to energies above 10 TeV in the future.

\subsection{Primary CR spectra}
\label{sec:protonhelium}

CR nuclei are typically divided into two classes,
the primary species accelerated by sources and secondary ones
produced due to fragmentation of primary CRs by collisions with the
interstellar medium (ISM) \citep{Strong:2007nh}. Highly abundant particles such as protons, helium, carbon, oxygen and iron nuclei are believed to be dominated by primary origin, whereas lithium, beryllium, boron and sub-iron nuclei belong to the secondary class. The spectra of primary CR nuclei reflect information about the particle acceleration in the source region, while the spectra of secondary ones and the secondary-to-primary flux ratios are crucial to understanding the propagation process of CRs. Thanks to its unique large geometrical factor and excellent charge resolution, DAMPE is capable of measuring the individual spectra of CR nuclei up to PeV scale.

\begin{figure*}[tbp]
  \centering
  \includegraphics[width=0.47\textwidth]{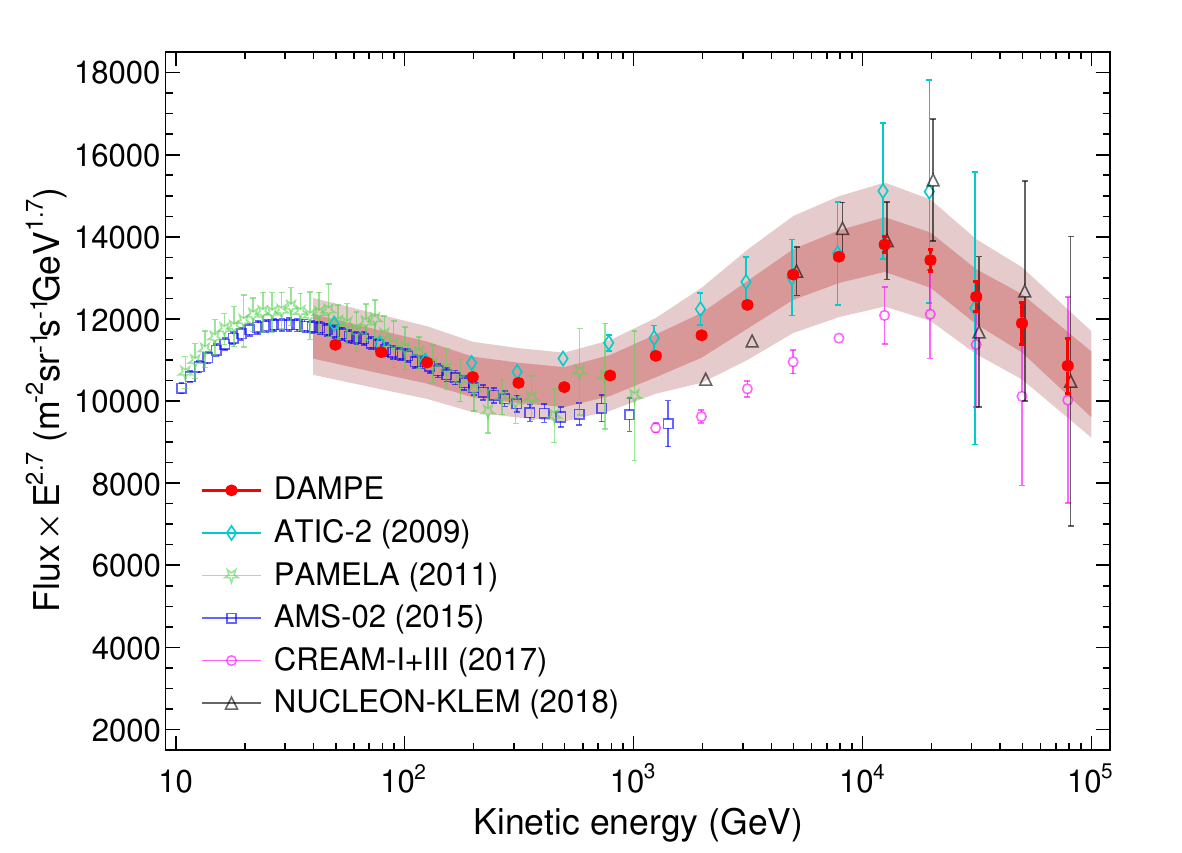}\hfill
  \includegraphics[width=0.47\textwidth]{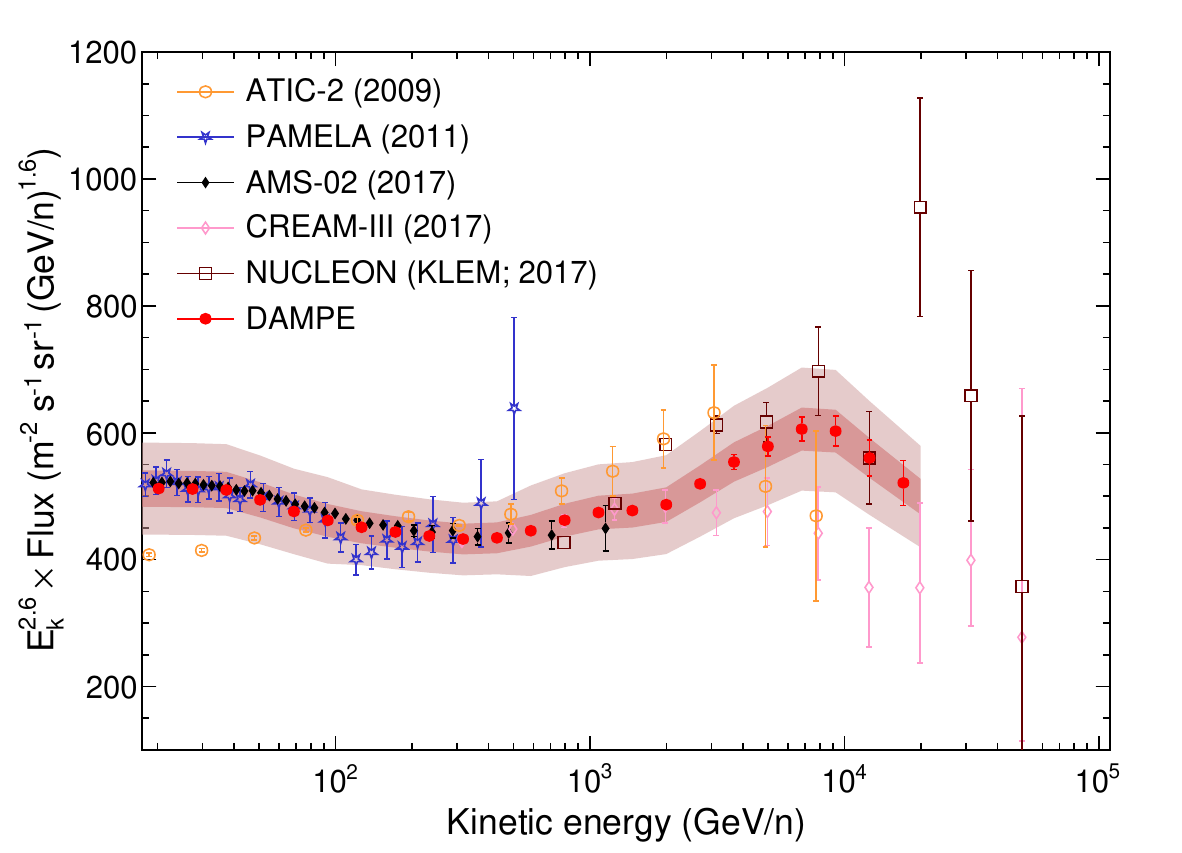}
  \caption{Cosmic-ray proton spectrum (left, Figure adapted from \citet{DAMPE:2019gys}) and helium spectrum (right, Figure adapted from \citet{DAMPE:2021gpb}) measured by DAMPE. Other measurements are from ATIC-2 \citep{Panov:2009iih}, PAMELA \citep{PAMELA:2011mvy}, AMS-02 \citep{AMS:2015tnn,AMS:2015azc}, CREAM \citep{Yoon:2017qjx} and NUCLEON \citep{Atkin:2018wsp}}
  \label{fig:phe}
\end{figure*}

Protons and helium nuclei are the two most abundant CR species, accounting for more than 99\% of the total cosmic rays. The excellent charge resolution enables DAMPE to have a powerful capability to identify protons and helium nuclei, and precisely measure their spectra respectively. With the first 30~months and 54~months of on-orbit data recorded by DAMPE, the CR proton spectrum from 40~GeV to 100~TeV \citep{DAMPE:2019gys} and the CR helium spectrum from 70~GeV to 80~TeV were precisely measured \citep{DAMPE:2021gpb}, as shown in Fig.~\ref{fig:phe}. The DAMPE measurements reveal that over a broad energy range the proton and helium spectra share similar features: a spectral hardening at energies of about a few hundred GeV and a spectral softening above 10 TeV. The softening above 10 TeV is clearly measured by the DAMPE experiment for the first time with high confidence (4.7$\sigma$ for protons and 4.3$\sigma$ for helium nuclei), with a break energy approximately proportional to the particle charge; this structure was subsequently further verified by the CALET experiment \citep{CALET:2022vro,CALET:2023nif}. This newly discovered spectral softening structure and its charge-dependent character are of important predictive significance for revealing the physical origin of the structure. DAMPE subsequently obtained a measurement of the CR proton+helium spectrum from 50~GeV to 500~TeV \citep{DAMPE:2023pjt}, which likewise reveals a structure of hardening followed by softening, with the softening significant at $>$6$\sigma$ standard deviations. 

\begin{figure}[tbp]
\centering
\includegraphics[width=0.47\textwidth]{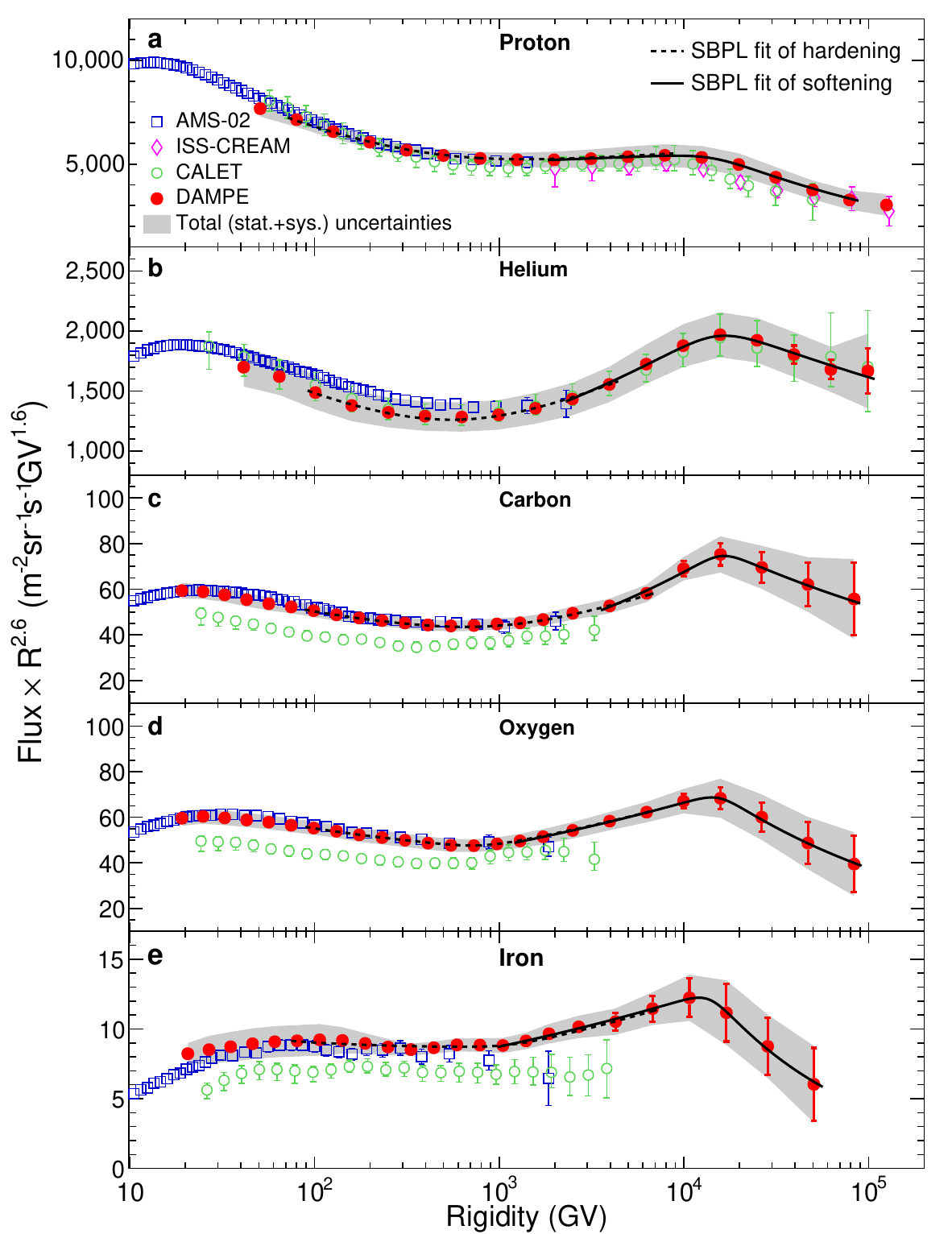}
\caption{Spectra of protons, helium, carbon, oxygen and iron nuclei measured by DAMPE (Figure adapted from \citet{DAMPE:2025opn}). Other measurements are from AMS-02 \citep{AMS:2021nhj, AMS:2021lxc}, CALET \citep{CALET:2022vro,CALET:2023nif,CALET:2020wyg,CALET:2021fks} and ISS-CREAM \citep{Choi:2022aht}.}
\label{fig:pHeCOFe}
\end{figure}

More recently, with the first 9~years of on-orbit data, DAMPE obtained measurements of the spectra of carbon, oxygen and iron nuclei from $\sim$20~GV to $\sim$100~TV (60~TV for iron), and updated the measurements of the proton and helium spectra \citep{DAMPE:2025opn}. The spectra of the five primary CR species are shown in Fig.~\ref{fig:pHeCOFe}. For the first time, distinct spectral softenings are directly detected in the spectra of carbon, oxygen and iron nuclei, by improving the precision considerably at high energies above several TV. Combined with the updated proton and helium spectra, the spectral softening appears universally at a rigidity of about 15 TV. A nuclei-mass-dependent softening is rejected at a confidence level of $>$99.999\% (4.4$\sigma$). 

\begin{figure*}[tbp]
\centering
\includegraphics[width=0.47\textwidth]{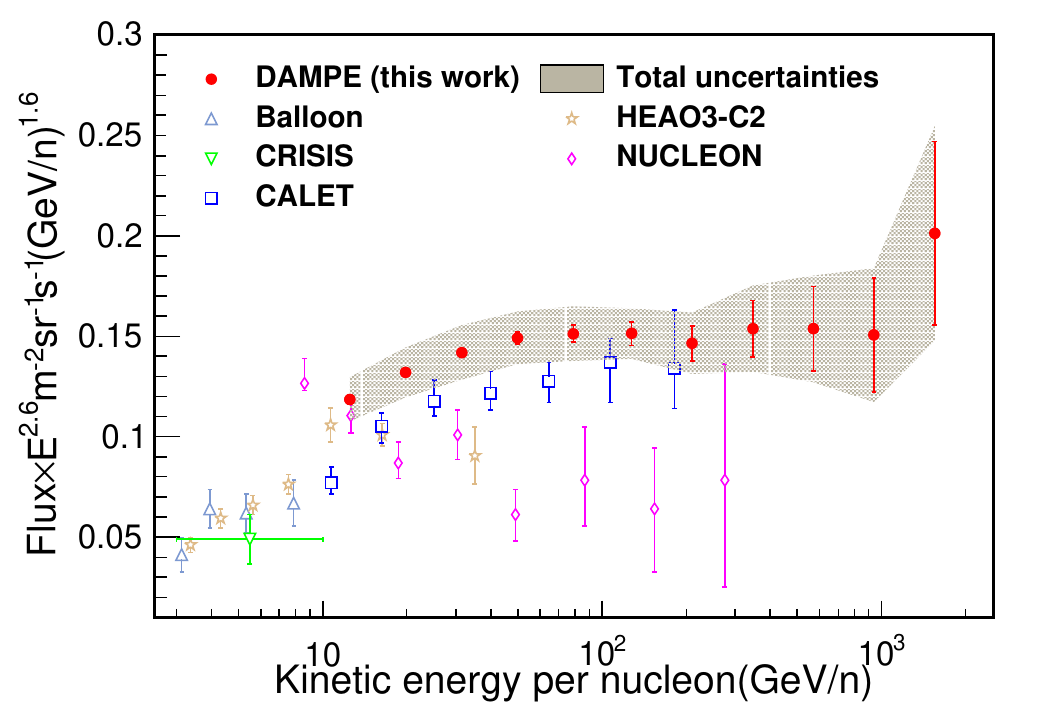}\hfill
\includegraphics[width=0.47\textwidth]{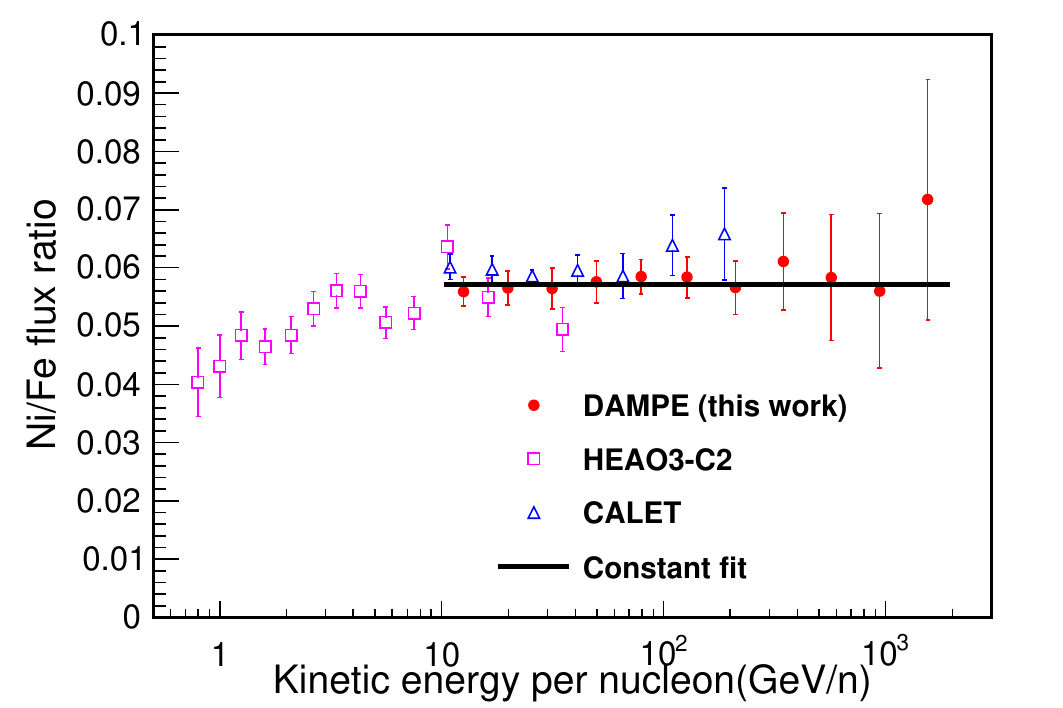}
\caption{The CR nickel spectrum as a function of kinetic energy per nucleon measured by DAMPE \citep{DAMPE:2025rzx}, compared with the measurements from CRISIS \citep{Young:1981}, HEAO-3-C2 \citep{Engelmann:1990zz},
NUCLEON \citep{Atkin:2019cab}, CALET \citep{CALET:2022ajb} and the earlier balloon-borne experiments \citep{Minagawa:1981} (left), and the nickel to iron flux ratios fit with a constant function (right) (Figures from \citet{DAMPE:2025rzx}).
}
\label{fig:ni}
\end{figure*}

The charge dependent spectral softenings of primary CRs indicate a common origin of the break due to acceleration or propagation effect. Combined with large-scale anisotropy measurements, the DAMPE measurements indicate a nearby CR accelerator, with the universal spectral softening marking its charge-dependent energy upper limit \citep{DAMPE:2025opn,Yuan:2025xqt}. The DAMPE observations provide the first direct experimental verification for the charge-dependent acceleration model, offering a key clue toward solving the longstanding mystery of CR origin.

For those CR species heavier than iron, DAMPE is still effective in identifying the individual elements, i.e. nickel nuclei, thanks to the excellent charge resolution of PSD. With 9~years of on-orbit data collected by DAMPE, the accurate measurement of the CR nickel spectrum was directly extended to TeV/n for the first time \citep{DAMPE:2025rzx}, as shown in Fig.~\ref{fig:ni} (left). The nickel spectrum is consistent with a single power law with spectral index -2.60$\pm$0.03 from 40~GeV/n to 1~TeV/n. In addition, the energy dependence of the nickel to iron flux ratio (Ni/Fe) is well described by a constant fit, which is structurally consistent with the CALET measurement \citep{CALET:2022ajb} in the same energy range, as shown in Fig.~\ref{fig:ni} (right). The constant Ni/Fe abundance ratio in CRs further indicates the similar origin, acceleration and propagation mechanism for both elements.

\subsection{Secondary boron spectrum}
\label{sec:secondary}

Boron nuclei in CRs are believed to be mainly produced by the fragmentation of heavier nuclei, such as carbon and oxygen, via collisions with the interstellar matter. Therefore, the spectra of boron nuclei and the B/C(O) flux ratios are very essential probes of the CR propagation. 

\begin{figure*}[tbp]
  \centering
  \includegraphics[width=0.47\textwidth]{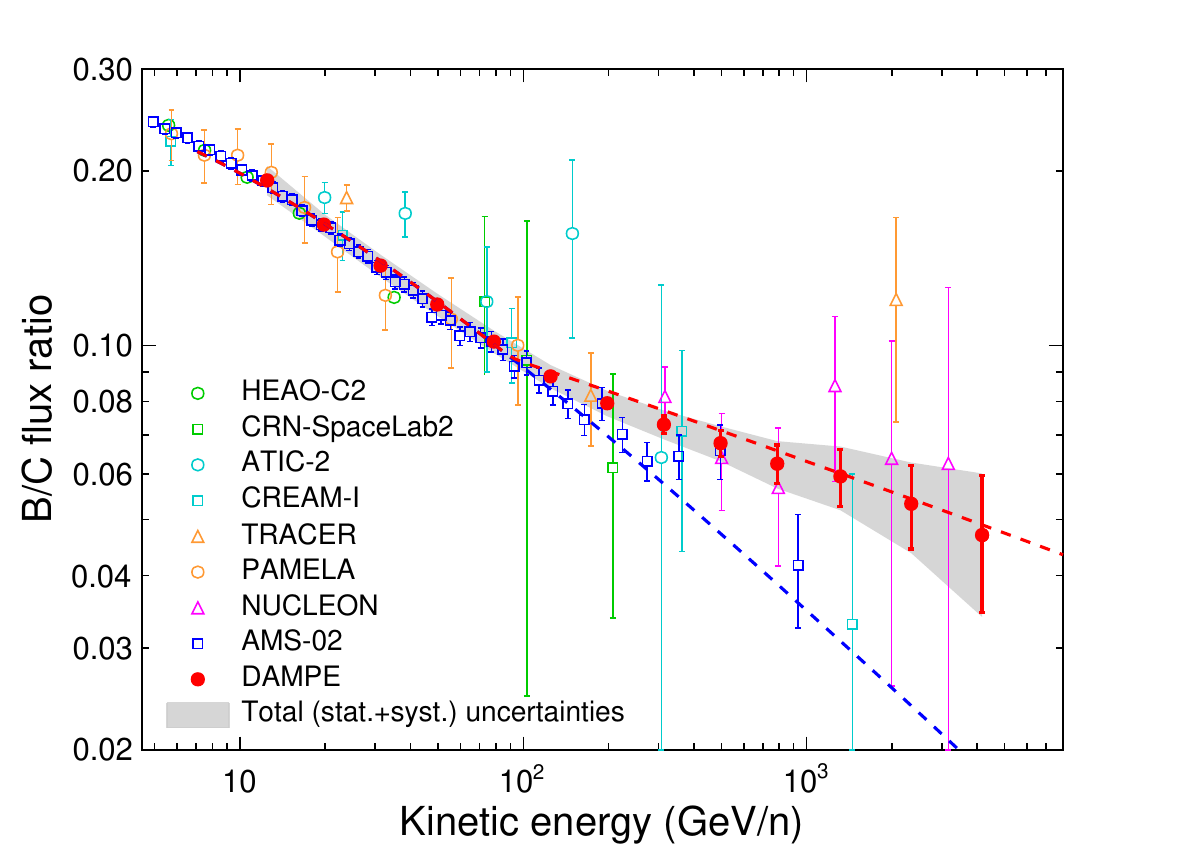}\hfill
  \includegraphics[width=0.47\textwidth]{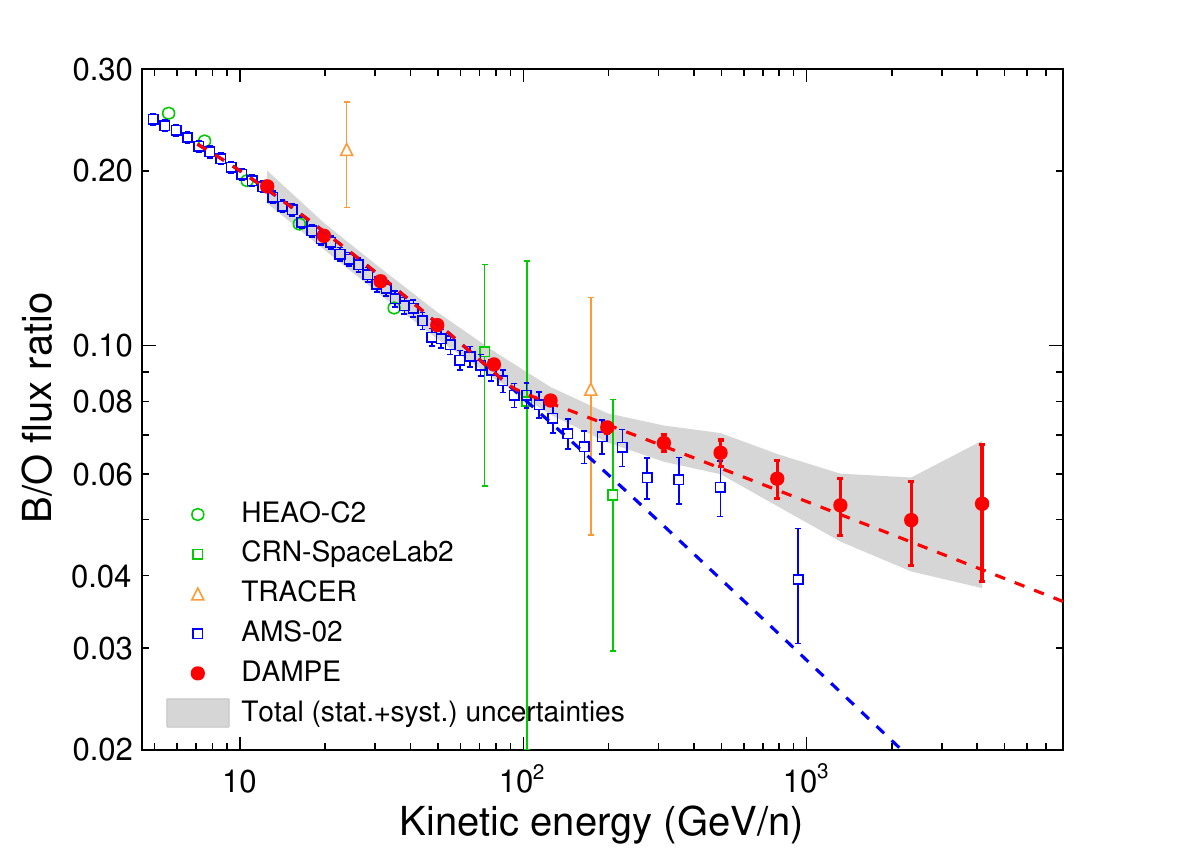}
  \caption{The B/C (left) and B/O (right) ratios measured by DAMPE (Figures from \citet{DAMPE:2022jgy}). Other measurements are from HEAO-C2 \citep{Engelmann:1990zz}, CRN-SpaceLab2 \citep{Swordy:1990}, ATIC-2 \citep{Panov:2007fe}, CREAM-I \citep{Ahn:2008my}, TRACER \citep{Obermeier:2011}, PAMELA \citep{Adriani:2014xoa}, NUCLEON \citep{Grebenyuk:2018jcb} and AMS-02 \citep{AMS:2021nhj}.}
  \label{fig:bcbo}
\end{figure*}

Based on the first 6~years of operation, DAMPE measured the B/C and B/O ratios precisely over the energy range from 10~GeV/n to 5.6~TeV/n \citep{DAMPE:2022jgy}, as shown in Fig.~\ref{fig:bcbo}. Compared with previous experimental results, the DAMPE measurements improve the accuracy above 100~GeV/n and extend to energies above 1~TeV/n for the first time. Most importantly, DAMPE detected a spectral hardening in both B/C and B/O ratios at about 100 GeV/n, which is inconsistent with the expectations of classical cosmic-ray propagation models. The most direct physical interpretation of this phenomenon is that the energy dependence of the cosmic-ray diffusion coefficient breaks at high energies \citep{Ma:2022iji}, so that cosmic rays propagate more slowly than expected in the higher-energy region, which increases their probability of interacting with the interstellar medium and producing secondary particles.

\begin{figure*}[tbp]
  \centering
  \includegraphics[width=0.47\textwidth]{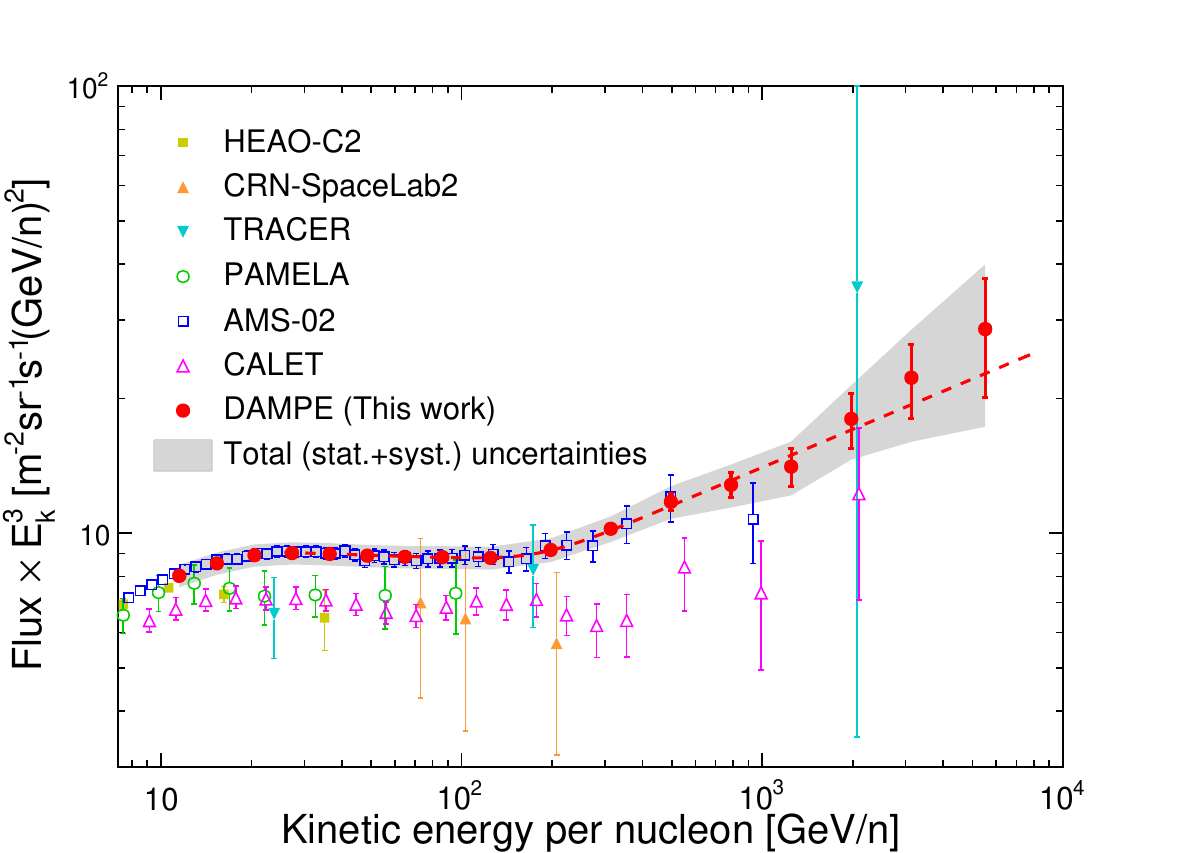}\hfill
  \includegraphics[width=0.47\textwidth]{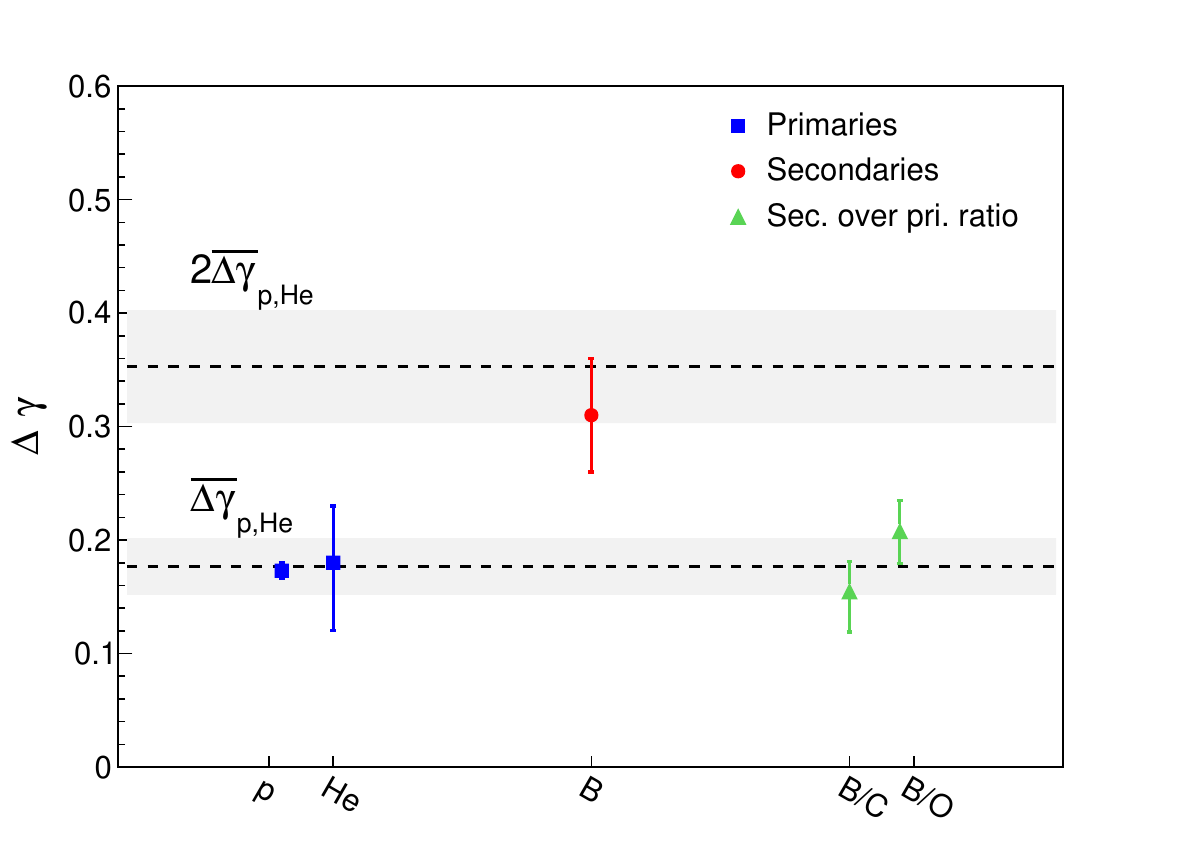}
  \caption{CR boron spectrum measured by DAMPE (left) and its spectral index change (right) (Figures from \citet{DAMPE:2024qwc}). Other measurements are from HEAO-C2 \citep{Engelmann:1990zz}, CRN-SpaceLab2 \citep{Swordy:1990}, TRACER \citep{Obermeier:2011}, PAMELA \citep{Adriani:2014xoa}, AMS-02 \citep{AMS:2023anq} and CALET \citep{CALET:2022dta}.}
  \label{fig:boron}
\end{figure*}

Subsequently, based on the 8 years of on-orbit data, DAMPE provided a precise measurement of the secondary boron spectrum from 10~GeV/n to 8~TeV/n \citep{DAMPE:2024qwc}, as shown in Fig.~\ref{fig:boron} (left). For the first time, DAMPE extended the precise measurement of the boron spectrum to energies above 1 TeV/n, and discovered a spectral hardening structure at about 200 GeV/n with a significance of 8 standard deviations. The amplitude of the spectral index change is about twice that of the primary particles proton and helium nuclei and of the B/C and B/O ratios, as shown in Fig.~\ref{fig:boron} (right). This indicates that the spectral hardening structure may originate from cosmic-ray propagation effects, as secondary boron nuclei undergo one more round of propagation and diffusion than primary particles do, and therefore exhibit a more severe spectral hardening structure.

\subsection{Gamma-ray line search}
\label{sec:line}

Thanks to the unprecedented high energy resolution, DAMPE is well suited for searching for monochromatic and sharp gamma-ray structures in the GeV–TeV range. With the first 5 years of on-orbit data collected by DAMPE, we carried out the search for gamma-ray lines from 10 to 300 GeV \citep{DAMPE:2021hsz}. To improve the sensitivity to line signals, two types of gamma-ray data sets, the LineSearch and BgoOnly data sets, are developed \citep{Xu:2021qok}.
The former includes events converted in the STK, which must traverse enough BGO layers to enhance energy resolution. The latter comprises events converted in the BGO calorimeter, where photons are reconstructed based on the shower axis in the BGO calorimeter.
This is the first time to take the BgoOnly data into the scientific analysis for the calorimeter-based gamma-ray observatories. 

\begin{figure*}[tbp]
\centering
\includegraphics[width=0.47\textwidth]{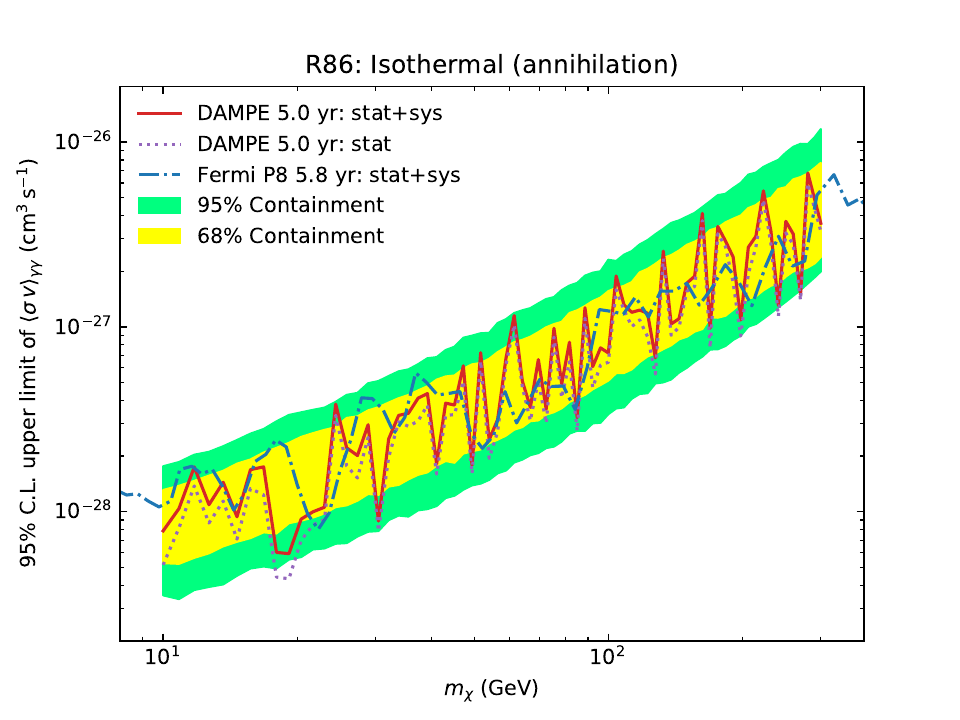}\hfill
\includegraphics[width=0.47\textwidth]{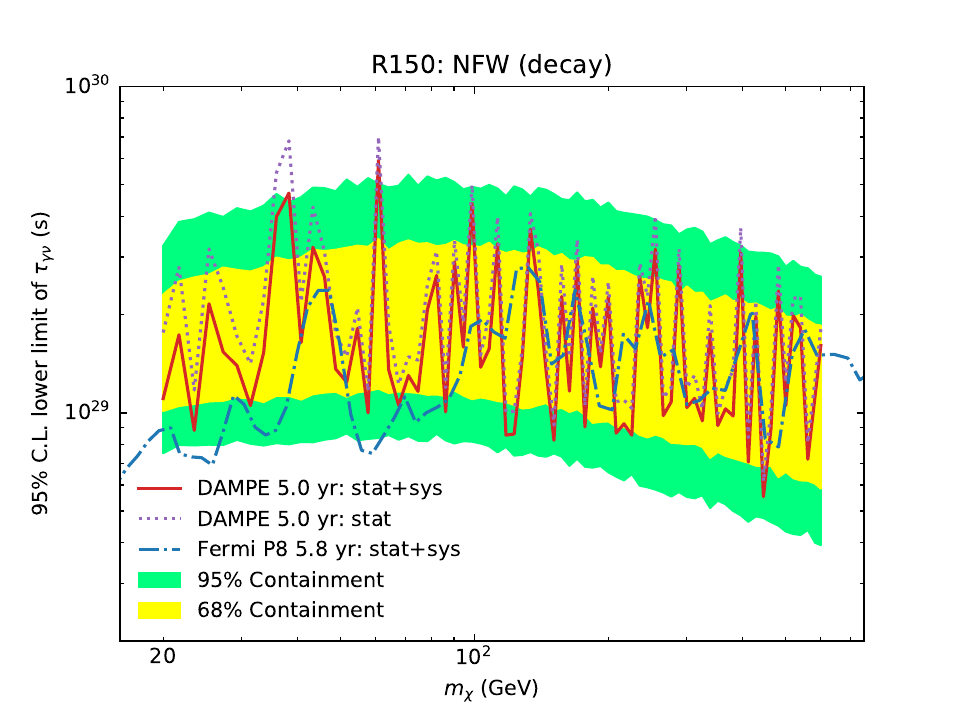}
\caption{The 95\% confidence level constraints for different DM density profiles (Figures from \citet{DAMPE:2021hsz}). The $\left< \sigma v\right>_{\gamma\gamma}$ upper limits of annihilating DM assuming the Isothermal profile (left), and the $\tau_{\gamma\nu}$ lower limit of decaying DM assuming the NFW profile (right). The red solid and purple dotted lines are the results with and without the systematic uncertainties respectively. The blue dot-dashed lines show the 5.8-year Fermi-LAT constraints \citep{Fermi-LAT:2015kyq}.
}
\label{fig:gammaline}
\end{figure*}

No line signals or candidates with TS value $\geq$ 9 are detected between 10 and 300 GeV in the Galaxy. The 95\% C.L. constraints on the annihilation cross section for $\chi\chi \to \gamma\gamma$ and the decay lifetime for $\chi \to \gamma\nu$, with proper addressing of the systematic uncertainties, are presented in Fig.~\ref{fig:gammaline} \citep{DAMPE:2021hsz}. The DAMPE constraints are comparable to the 5.8-year results of Fermi-LAT \citep{Fermi-LAT:2015kyq} thanks to the better energy resolution and the smaller influence of the systematic uncertainties. For the decaying DM, the DAMPE lower limits on the decay lifetime are stronger for DM with mass $\leq$ 100~GeV by a factor of $\sim$ 2. In view of the fact that the DAMPE data set is about ten times smaller than that of Fermi-LAT, our findings demonstrate the potential of high-energy-resolution observations on dark matter detection. As anticipated, combining the high-energy resolution DAMPE dataset with the large photon statistics of  Fermi Large Area Telescope (Fermi-LAT) significantly enhances the sensitivity of dark matter line searches \citep{2023PhRvD.108f3015C}.

\section{Summary and future prospects}
\label{sec:outlook}
Since its launch at the end of 2015, the DAMPE instrument has maintained an excellent performance and operated stably in space for more than 10 years. With the help of high energy resolution and a relatively large effective detection area, the DAMPE experiment has achieved a series of important advances in the direct measurement of CR energy spectra, precisely determining the energy spectra of multiple cosmic-ray components over a broad energy range and discovering new spectral break structures, which provides key data for understanding the mechanisms of the origin, acceleration and propagation of Galactic CRs. Beyond its routine scientific goals, DAMPE has made significant progress in expanded research areas, including observations of solar flare activity \citep{DAMPE:2021qet,DAMPE:2026xwf} and searches for fractional charge particles in space \citep{DAMPE:2022zzr,DAMPE:2026qrp}. At present, the DAMPE satellite continues to operate stably in orbit with good detector performance, and analyses of the all-electron spectrum extended to 10~TeV and of the spectra of other CR nuclei are in progress. Also, a joint measurement of the cosmic-ray spectrum by DAMPE and LHAASO \citep{LHAASO:2025byy,LHAASO:2025mlf} is under way, with precise measurements of the Galactic cosmic-ray proton spectrum across a very wide energy range, providing key clues to the origin of Galactic cosmic rays.

In addition, China is actively advancing preliminary research for the next generation of space-based direct high-energy particle detection experiments, including such important projects as the High Energy cosmic-Radiation Detection facility (HERD; \citet{HERD:2014bpk}), planned for deployment on China's Space Station, and the Very Large Area gamma-ray Space Telescope (VLAST; \citet{2022AcASn..63...27F}) \citep{Pan:2024adp}. With these advanced observational facilities, it is expected that precise measurements of the spectral structures and anisotropy characteristics of various cosmic-ray components over an ultra-wide energy range can be achieved, ultimately providing decisive evidence for solving the mystery of the origin of cosmic rays and the nature of dark matter (see Fig.4 of \citet{Fan:2024rcr} for an example).

\begin{acknowledgments}
The DAMPE mission was funded by the strategic priority science and
technology projects in space science of the Chinese Academy of Sciences (CAS). This work was mainly supported by the National Natural Science Foundation of China (No. 12588101), the National Key Research and Development Program of China (No. 2022YFF0503304), and the Strategic Priority Program on Space Science of CAS (No. E02212A02S).

\end{acknowledgments}

\bibliographystyle{astronomycomm}
\makeatletter
\def \@biblabel#1{}
\bibliography{dampe}

\end{document}